\documentclass[prb,floatfix,twocolumn,showpacs,amsmath,amssymb]{revtex4}
\usepackage{graphicx}
\usepackage{dcolumn}
\usepackage{bm}
\begin{document}

\title{Spin-liquid and magnetic phases in the anisotropic triangular lattice:\\ 
the case of $\kappa$-(ET)$_2$X}
\author{Luca F. Tocchio,$^{1}$ Alberto Parola,$^{2}$, Claudius Gros,$^{1}$, 
and Federico Becca$^{3}$}
\affiliation{
$^{1}$ Institute for Theoretical Physics, Goethe-University
Frankfurt, Max-von-Laue-Stra{\ss}e 1, D-60438 Frankfurt am Main, Germany \\
$^{2}$ Dipartimento di Fisica e Matematica, Universit\`a dell'Insubria,
Via Valleggio 11, I-22100 Como, Italy \\
$^{3}$ CNR-INFM-Democritos National Simulation Centre and International School
for Advanced Studies (SISSA), Via Beirut 2, I-34151 Trieste, Italy
}

\date{\today} 

\begin{abstract}
The two-dimensional Hubbard model on the anisotropic triangular lattice, with
two different hopping amplitudes $t$ and $t^\prime$, is relevant to describe 
the low-energy physics of $\kappa$-(ET)$_2$X, a family of organic salts.
The ground-state properties of this model are studied by using Monte Carlo 
techniques, on the basis of a recent definition of backflow correlations 
for strongly-correlated lattice systems. The results show that there is no 
magnetic order for reasonably large values of the electron-electron interaction
$U$ and frustrating ratio $t^\prime/t = 0.85$, suitable to describe the 
non-magnetic compound with X=Cu$_2$(CN)$_3$. On the contrary, N\'eel order 
takes place for weaker frustrations, i.e., $t^\prime/t \sim 0.4 \div 0.6$,
suitable for materials with X=Cu$_2$(SCN)$_2$, Cu[N(CN)$_2$]Cl, or 
Cu[N(CN)$_2$]Br.
\end{abstract}


\maketitle

\section{Introduction}
Organic charge-transfer salts show a wide variety of quantum phases and 
represent prominent examples to study correlation effects in low-dimensional 
systems. The most celebrated case is given by the TTF-TCNQ salt that has been
primarily regarded as a prototype for testing theories of one-dimensional 
conductors.~\cite{solyom} Organic salts may also form crystals in two and 
three dimensions, and, in this respect, an increasing attention has been 
devoted to a particular family denoted by $\kappa$-(ET)$_2$X, whose 
building block is the so-called BEDT-TTF (or ET) molecule and X is a monovalent
anion.~\cite{kanoda} Here, strongly dimerized ET molecules are arranged in a 
two-dimensional triangular lattice. Each dimer has a charge state with one 
hole and therefore the conducting band is half filled. A sizable effective 
Coulomb repulsion is felt by two holes on the same dimer.
A huge variety of phases have been found (by varying temperature, 
pressure or the nature of the anion X), ranging from correlated (bad) metals 
with superconductivity at low temperatures, to Mott insulators with magnetic 
order.~\cite{kanodamag,elsinger,lefebvre,limelette} 
Interestingly, by acting with hydrostatic pressure, metal-insulator 
transitions have been observed,~\cite{kanodamott1,kanodamott2} with the 
remarkable possibility to stabilize a non-magnetic Mott insulating phase in 
$\kappa$-(ET)$_2$Cu$_2$(CN)$_3$.~\cite{kanodaliquid} In this material, there 
is no evidence of magnetic order down to $T \simeq 30 mK$, which is four 
orders of magnitude lower than the estimate of the super-exchange coupling 
$J \simeq 250 K$. 

It has been argued that $\kappa$-(ET)$_2$X compounds can be described by a 
single-band Hubbard model on the anisotropic triangular lattice,~\cite{kino} 
where chains described by an hopping $t^\prime$ are coupled together with 
zig-zag hoppings $t$, see Fig.~\ref{fig:lattice}. An on-site repulsive term 
$U$ is also present in the Hamiltonian. However, a realistic estimate of these 
microscopic parameters is not exempt from complications. 
Indeed, the values obtained some time ago by extended H\"uckel band structure 
calculations~\cite{mckenzie} have been put in doubt by two recent ab-initio 
calculations, based upon local-density approximation (LDA) and generalized 
gradient approximation (GGA).~\cite{nakamura,valenti} 
Interestingly, the new results suggest that these organic salts are less 
frustrated than previously assumed, and that $t^\prime/t$ is smaller than one. 
Indeed, the frustrating ratio is $t^\prime/t \sim 0.8$ for 
$\kappa$-(ET)$_2$Cu$_2$(CN)$_3$ and $t^\prime/t \sim 0.6$ for 
$\kappa$-(ET)$_2$Cu$_2$(SCN)$_2$.~\cite{nakamura,valenti} 
Other materials, with X=Cu[N(CN)$_2$]Cl or Cu[N(CN)$_2$]Br, have a 
substantially smaller frustrating ratio, i.e., 
$t^\prime/t \sim 0.4$.~\cite{valenti}
Unfortunately, an accurate determination of the correlation energy is rather 
difficult and these two calculations give a considerably different estimation
of the Coulomb repulsion, namely $U/t \sim 12 \div 15$ 
(Ref.~\onlinecite{nakamura}) and $U/t \sim 5 \div 7$ 
(Ref.~\onlinecite{valenti}).

Here, we apply our improved Monte Carlo calculations, based
upon the recently introduced backflow wave function~\cite{tocchio} 
in order to analyze the possibility of having a non-magnetic 
insulator for large enough frustration and interaction strength.

The paper is organized as follow: in section~\ref{sec:model}, we introduce the 
Hamiltonian; in section~\ref{sec:approach}, we describe our variational wave
function; in section~\ref{sec:results}, we present our numerical results and,
finally, in section~\ref{sec:conc} we draw the conclusions.

\section{Model}\label{sec:model}
\begin{figure}
\includegraphics[width=\columnwidth]{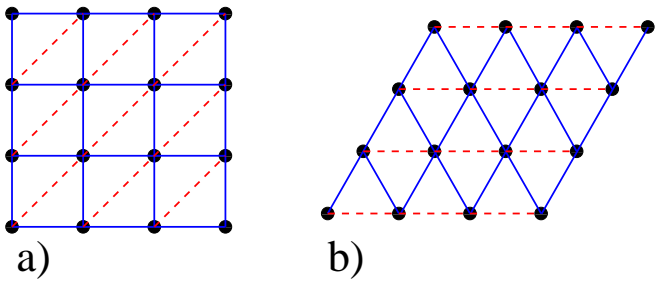}
\caption{\label{fig:lattice}
Illustration of the lattice in the square topology (a) used in this work
and in the equivalent triangular one (b). Solid and dashed lines indicate 
hopping amplitudes $t$ and $t^\prime$, respectively.}
\end{figure}

We consider the Hubbard model described by
\begin{equation}\label{hubbard}
{\cal H}=-\sum_{i,j,\sigma} t_{ij} c^{\dagger}_{i,\sigma} c_{j,\sigma} + H.c.
+U \sum_{i} n_{i,\uparrow} n_{i,\downarrow},
\end{equation}
where $c^{\dagger}_{i,\sigma} (c_{i,\sigma})$ creates (destroys) an electron 
with spin $\sigma$ on site $i$, 
$n_{i,\sigma}=c^{\dagger}_{i,\sigma}c_{i,\sigma}$, $t_{ij}$ is the hopping
amplitude and $U$ is the on-site Coulomb repulsion. In this work, we focus 
our attention on the half-filled case with $N$ electrons on $N$ sites and 
consider a square lattice with a nearest-neighbor hopping $t$, along the
$(1,0)$ and $(0,1)$ directions, and a further next-nearest-neighbor hopping 
$t^\prime$ along $(1,1)$; this choice of the hopping amplitudes is 
topologically equivalent to the anisotropic triangular lattice,~\cite{notet} 
see Fig.~\ref{fig:lattice}.
In the last years, an intense effort has been devoted to this problem 
by use of a large variety of methods, including exact 
diagonalization,~\cite{clay} path-integral renormalization 
group,~\cite{imada} variational Monte Carlo,~\cite{watanabe1,watanabe2,zhang} 
cluster dynamical mean field theory,~\cite{kyung,ohashi} and dual 
Fermions.~\cite{lee} 
All these methods give rather different outcomes and there are huge 
discrepancies on the phase boundaries and, most importantly, on the expected 
nature of the non-magnetic insulator. The aim of this work is to clarify the 
ground-state properties for two values of $t^\prime/t=0.6$ and $0.85$, 
relevant for materials with X=Cu$_2$(SCN)$_2$ and Cu$_2$(CN)$_3$, respectively.

\begin{figure}
\includegraphics[width=\columnwidth]{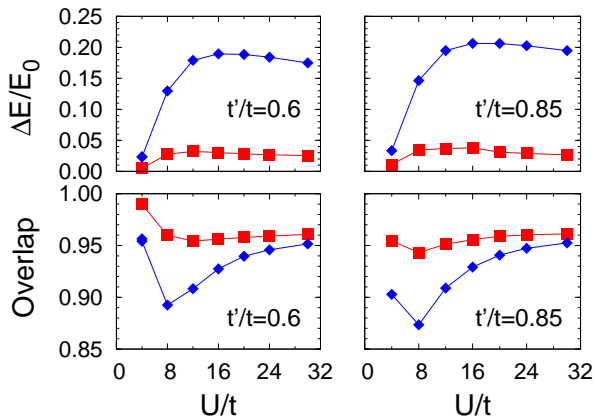}
\caption{\label{fig:accuracy}
(Color online) Results for 18 electrons on 18 sites as a function of $U/t$.
Upper panels: Accuracy of energy $\Delta E=(E_0-E_v)$, $E_v$ and $E_0$ being
the variational and the exact values, respectively. Lower panels: Overlap
between the exact ground state and the variational BCS wave functions. The
state without (with) backflow correlations is denoted by diamonds (squares).}
\end{figure}

\section{Variational approach}\label{sec:approach}
A variational wave function for an insulator with antiferromagnetic (AF) order 
can be constructed by considering the ground state $|AF\rangle$ of a 
mean-field Hamiltonian containing a band contribution and a magnetic term:
\begin{equation}
{\cal H}_{AF} = \sum_{q,\sigma} \epsilon_q c^{\dagger}_{q,\sigma} c_{q,\sigma} 
+ \Delta_{AF} \sum_i {\bf n}_i \cdot {\bf S}_i,
\end{equation}
where ${\bf n}_i$ is a unitary vector that depends upon the lattice site 
$i$ and ${\bf S}_i=(S_i^x,S_i^y,S_i^z)$ is the spin operator. Moreover,
$\epsilon_q = -2t(\cos q_x+\cos q_y) -2t_d \cos(q_x+q_y)$ is a variational 
band term: $t$ gives the energy scale and $t_d$ can be optimized to minimize
the variational energy. In order to correctly describe spin fluctuations 
orthogonal to the plane where the magnetic order lies, we take ${\bf n}_i$ 
in the $x{-}y$ plane and we include a spin Jastrow factor 
${\cal J}_s=\exp [-\frac{1}{2} \sum_{i,j} u_{i,j} S_i^z S_j^z ]$
in the wave function.~\cite{becca} Another density Jastrow factor
${\cal J}=\exp [-\frac{1}{2} \sum_{i,j} v_{i,j} n_i n_j ]$ (that includes the
on-site Gutzwiller term) is considered to adjust electron correlations.
In summary, the correlated wave function is defined by
\begin{equation}
|\Psi_{AF}\rangle = {\cal J}_s {\cal J} |AF\rangle.
\end{equation}
Notice that, in this case, the variational state has not a definite total spin,
which is suitable for a magnetically ordered phase. In fact, both $|AF\rangle$ 
and the spin Jastrow factor ${\cal J}_s$ break the SU(2) symmetry. 

On the other hand, superconducting or metallic phases can be constructed by 
considering the ground state $|BCS\rangle$ of a superconducting 
Bardeen-Cooper-Schrieffer (BCS) Hamiltonian with both band and pairing 
contributions,~\cite{gros,zhang88}
\begin{equation}
{\cal H}_{BCS} = \sum_{q,\sigma} \epsilon_q c^{\dagger}_{q,\sigma} c_{q,\sigma}
+ \sum_{q} \Delta_q 
c^{\dagger}_{q,\uparrow} c^{\dagger}_{-q,\downarrow} + H.c.,
\end{equation}
here the band term may also contain a variational chemical potential $\mu$,
since the BCS Hamiltonian does not conserve the particle number, i.e.,
$\epsilon_q = -2t(\cos q_x+\cos q_y) -2t_d \cos(q_x+q_y) -\mu$. In this case,
$t_d$ and $\mu$ can be varied to optimize the variational wave function.
The full correlated state is given by
\begin{equation}
|\Psi_{BCS}\rangle = {\cal J} |BCS\rangle,
\end{equation}
in this case, no spin Jastrow is considered, in order to have a perfect singlet
state, suitable for a non-magnetic phase. Notably, within this kind of wave 
function, it is possible to obtain a pure (i.e., non-magnetic) Mott insulator 
just by considering a sufficiently strong Jastrow factor, i.e., 
$v_q \sim 1/q^2$ ($v_q$ being the Fourier transform of 
$v_{i,j}$).~\cite{capello} 

As we recently demonstrated,~\cite{tocchio} the projected BCS state is
not sufficiently accurate for Hubbard-type models, especially in the important 
strong-coupling regime, i.e., for $U/t \gtrsim 10$, where the super-exchange 
energy scale $J=4t^2/U$ is not correctly reproduced. One efficient way to 
overcome this problem is to consider backflow correlations,~\cite{tocchio} 
that modify the single-particle orbitals,~\cite{noteph} in the same spirit 
of what was put forward long-time ago by Feynman and Cohen.~\cite{feynman} 

\begin{figure}
\includegraphics[width=\columnwidth]{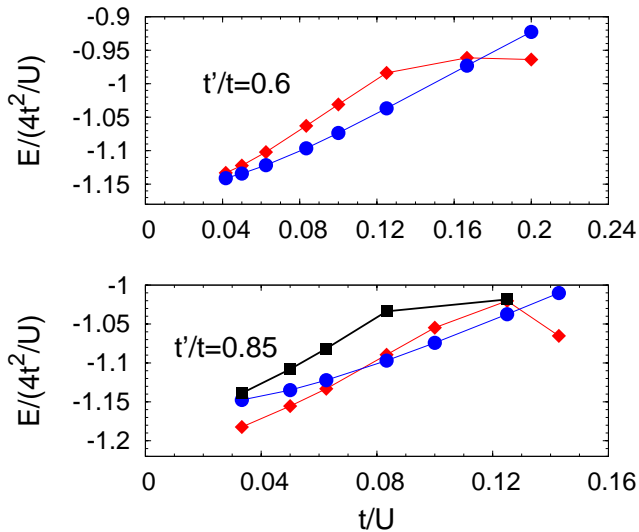}
\caption{\label{fig:energy}
(Color online) Variational energies per site (in unit of $J=4t^2/U$) for the
BCS state with a density Jastrow factor (diamonds) and for the AF wave function
with both density and spin Jastrow terms (circles); The correlated Fermi gas 
with Jastrow factor is also reported for $t^\prime/t=0.85$ (squares). 
All states have backflow correlations and results are for 100 sites.}
\end{figure}

Following Ref.~\onlinecite{tocchio}, we consider a general definition of
the new ``orbitals'' by taking all the possible virtual hoppings of the 
electrons:
\begin{eqnarray}\label{backlattice2}
&&\phi_q^b({\bf r}_{i,\sigma}) \equiv \epsilon \phi_q({\bf r}_{i,\sigma})
+ \eta_1 \sum_j t_{ij} D_i H_j \phi_q({\bf r}_{j,\sigma})
+ \nonumber \\
&& \eta_2 \sum_j t_{ij} n_{i,\sigma} h_{i,-\sigma}
n_{j,-\sigma} h_{j,\sigma} \phi_q({\bf r}_{j,\sigma}) + \nonumber \\
&& \eta_3 \sum_j t_{ij} \left( D_i n_{j,-\sigma} h_{j,\sigma}
+n_{i,\sigma} h_{i,-\sigma} H_j \right) \phi_q({\bf r}_{j,\sigma}),
\end{eqnarray}
where we used the notation
$\phi_q({\bf r}_{i,\sigma})= \langle 0|c_{i,\sigma}|\phi_q\rangle$, being 
$|\phi_q\rangle$ the eigenstates of the mean-field Hamiltonian,
$D_i=n_{i,\uparrow}n_{i,\downarrow}$, $H_i=h_{i,\uparrow}h_{i,\downarrow}$, 
with $h_{i,\sigma}=(1-n_{i,\sigma})$.
$\epsilon$, $\eta_1$, $\eta_2$, and $\eta_3$ are variational parameters.
As a consequence, already the determinant part of the wave function includes 
correlation effects. The backflow corrections of Eq.~(\ref{backlattice2}) 
(in particular the $\eta_1$ term) make it possible to mimic the effect of the 
virtual hopping, which leads to the super-exchange mechanism. 
All the parameters of the wave function can be optimized by using the method of 
Ref.~\onlinecite{yunoki}. 

Finally, the accuracy of the variational calculations can be assessed by using 
Lanczos diagonalizations on small lattices and Green's function Monte Carlo 
within the so-called fixed-node approximation,~\cite{ceperleyfn} which 
gives accurate (but approximate) results on large systems. A detailed 
description of the fixed-node approximation can be found 
in Ref.~\onlinecite{lugas}. In brief, starting from the original Hamiltonian 
${\cal H}$, we define an effective Hamiltonian by adding a perturbation $O$:
\begin{equation}\label{fnham}
{\cal H}_{eff} = {\cal H} + O.
\end{equation}
The operator $O$ is defined through its matrix elements and depends upon
a given {\it guiding function} $|\Psi \rangle$, that is for instance the 
variational state itself
\begin{equation}
O_{x^\prime,x} = \left \{
\begin{array}{ll}
-{\cal H}_{x^\prime,x} & {\rm if} \; x\prime \neq x \; {\rm and}\; s_{x^\prime,x} > 0 \nonumber \\
0 & {\rm if} \; x\prime \neq x \; {\rm and}\; s_{x^\prime,x} < 0 \nonumber \\
\sum_{y:s_{y,x}>0} {\cal H}_{y,x} \frac{\Psi_y}{\Psi_x} & {\rm for} \; x^\prime=
x,
\end{array}
\right .
\end{equation}
where $\Psi_x = \langle x|\Psi \rangle$ and $s_{x^\prime,x} = \Psi_{x^\prime} {\cal H}_{
x^\prime,x} \Psi_x$.
Notice that the above operator annihilates the guiding function, namely
$O |\Psi \rangle=0$. Therefore, whenever the guiding function is close
to the exact ground state of ${\cal H}$, the perturbation $O$ is expected to be 
small and the effective Hamiltonian becomes very close to the original one.
 
\section{Results}\label{sec:results}
By allowing the most general singlet and complex BCS pairing in the state
without backflow terms, we find that this quantity has $d_{x^2-y^2}$ symmetry 
up to $t^\prime \sim t$, namely the best (nearest-neighbor) pairing function is
$\Delta_q= 2 \Delta (\cos q_x - \cos q_y)$, in agreement with previous 
results.~\cite{powell,zhang2,nandini,powell2}
Therefore, within our improved backflow wave function, we only considered 
a real BCS pairing. We mention that $\Delta$ is very small (especially in the
presence of backflow correlations) in the conducting phase, 
see table~\ref{tab:delta}, and it becomes sizable only in the regime where the 
magnetic solution prevails over the BCS state. In this regard, we do not find 
a clear signature of superconductivity close to the metal-insulator transition, 
as suggested in Ref.~\onlinecite{nandini}. 
We also stress that, once the backflow correlations are considered, there is no
energy gain by allowing a translational symmetry breaking (e.g., by considering
a $2 \times 1$ unit cell in the BCS Hamiltonian, suitable for dimerized states)
and the $d_{x^2-y^2}$ solution has always a lower energy than dimerized states.
Finally, we find that the variational band term of the BCS Hamiltonian
$\epsilon_q= -2t(\cos q_x + \cos q_y) -2t_d \cos(q_x+q_y) -\mu$
has $t_d \simeq 0$ for most of the cases considered, except for small $U/t$, 
inside the conducting phase, where a finite $t_d$ can be stabilized. 

\begin{table}
\caption{\label{tab:delta}
BCS pairing $\Delta$ for various $U/t$ in the metallic region for two sizes
of the lattice: $N=100$ (third column) and $N=196$ (fourth column).
Notice that for $U/t=8$ and $t^\prime/t=0.85$ and for $U/t=6$ and 
$t^\prime/t=0.6$ the BCS wave function is still metallic but the AF one 
(insulating) has a lower variational energy.}
\begin{tabular}{cccc}
\hline
$U/t$ & $t^\prime/t$ & $\Delta/t$ & $\Delta/t$ \\
\hline \hline
6     & 0.85         & 0.026(1) & 0.018(1) \\
7     & 0.85         & 0.051(1) & 0.025(1) \\
8     & 0.85         & 0.161(1) & 0.037(1) \\
\hline \hline
4     & 0.6          & 0.013(1) & 0.005(1) \\
5     & 0.6          & 0.027(1) & 0.008(1) \\
6     & 0.6          & 0.056(1) & 0.019(1) \\
\hline \hline
\end{tabular}
\end{table}

As far as the magnetic wave function is concerned, both Hartree-Fock and 
fixed-node calculations give a clear indication that spin-spin correlations 
remain commensurate at $Q=(\pi,\pi)$ for $t^\prime/t \lesssim 0.9$. Therefore, 
we use an AF wave function having N\'eel order with pitch vector $Q=(\pi,\pi)$
and we do not consider the implementation of a generic magnetic state with
incommensurate order. Moreover, we verified that, for 
$t^\prime/t \lesssim 0.9$, this AF state has a lower energy with respect to 
the AF state with 120$^\circ$ order, suitable for $t^\prime=t$.

\begin{figure}
\includegraphics[width=\columnwidth]{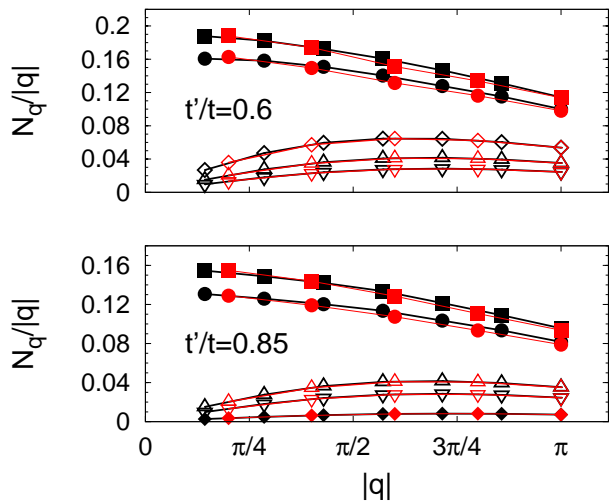}
\caption{\label{fig:nq}
(Color online) Variational results for the density-density correlations $N_q$ 
divided by $|q|$, along the $(1,0)$ direction, for 100 (red symbols) and 196 
(black symbols) sites. Full (empty) symbols refer to the BCS (AF) wave 
function. Upper panel: from top to bottom, $U/t=4$, $5$, $6$, $8$, and $10$. 
Lower panel: from top to bottom, $U/t=6$, $7$, $8$, $10$, and $20$.}
\end{figure}

\subsection{Quality of the variational states}

In Fig.~\ref{fig:accuracy}, we show the accuracy of the BCS variational state
and its overlap with the exact ground state in a small lattice with 18 sites
(which is tilted by 45 degrees). We report two cases with $t^\prime/t=0.6$ 
and $0.85$ and different values of $U/t$. As in the case of the frustrated 
square lattice studied in Ref.~\onlinecite{tocchio}, the backflow terms highly 
improve the quality of the variational wave function that remains very accurate
even for large correlation, i.e., up to $U/t \sim 30$. We would like to mention
that, for this small cluster, the AF state has a slightly lower energy than 
the BCS one for both $t^\prime/t=0.6$ and $0.85$. For $t^\prime/t=0.6$, the AF 
state has also a better overlap with the exact ground state $|\Psi_0\rangle$ 
(e.g., $\langle \Psi_0|\Psi_{AF}\rangle = 0.962$ for $U/t=20$) than the BCS 
state (e.g., $\langle \Psi_0|\Psi_{BCS}\rangle = 0.958$), while it has a 
substantially lower overlap for $t^\prime/t=0.85$ (e.g., 
$\langle \Psi_0|\Psi_{AF}\rangle = 0.904$ against 
$\langle \Psi_0|\Psi_{BCS}\rangle = 0.959$).

The accuracy of the variational state remains very high also for large 
systems, where the backflow corrections give a sizable and size-consistent
improvement. In Fig.~\ref{fig:energy}, we report the energy per site as a 
function of $U/t$ for both $t^\prime/t=0.6$ and $0.85$ for $N=10 \times 10$ 
(see also table~\ref{tab:energies}). 

\begin{table}
\caption{\label{tab:energies}
Our best energies per site for $N=100$: pure variational $E_{vmc}$ and
fixed-node $E_{fn}$ (still variational) results are reported.}
\begin{tabular}{cccc}
\hline
$U/t$ & $t^\prime/t$ & $E_{vmc}/t$  & $E_{fn}/t$  \\
\hline \hline
4     & 0.85         & -1.03029(2)  & -1.0315(1) \\
8     & 0.85         & -0.51876(5)  & -0.5238(1) \\
12    & 0.85         & -0.36569(5)  & -0.3764(1) \\
16    & 0.85         & -0.2834(1)   & -0.2910(1) \\
20    & 0.85         & -0.2311(1)   & -0.2364(1) \\
\hline \hline
4     & 0.6          & -0.92356(2)  & -0.9251(1) \\
8     & 0.6          & -0.51837(3)  & -0.5228(1) \\
12    & 0.6          & -0.36550(3)  & -0.3689(1) \\
16    & 0.6          & -0.28041(3)  & -0.2833(1) \\
20    & 0.6          & -0.22685(3)  & -0.2291(1) \\
\hline \hline
\end{tabular}
\end{table}

\subsection{Metal-insulator transition}

The metal-insulator transition can be detected by a direct inspection of the 
static density-density correlations
\begin{equation}
N_q = \frac{1}{N} \sum_{j,l} e^{i q (R_j-R_l)} \langle n_j n_l \rangle.
\end{equation}
In fact, this quantity makes it possible to discriminate between gapless 
(conducting) and gapped (insulating) phases: a linear behavior $N_q \sim |q|$ 
for $|q| \to 0$ is typical of a conducting phase, whereas a quadratic behavior
$N_q \sim q^2$ can be associated to an insulating character.~\cite{capello} 
The results presented in Fig.~\ref{fig:nq} indicate that a metal-insulator 
transition takes place by increasing $U/t$ and it can be placed at 
$U_c^{\rm MIT}/t= (5.5 \pm 0.5)$ and $(7.5 \pm 0.5)$ for $t^\prime/t=0.6$ 
and $0.85$, respectively.
The transition is first order, with a small jump in the linear coefficient 
of $N_q$ for small momenta. In fact, for small $U/t$, the best wave function 
is the BCS one (with small superconducting pairing), whereas, by increasing 
the interaction, the AF one prevails, thus inducing a metal-insulator 
transition, see Fig.~\ref{fig:energy}.

\begin{figure}
\includegraphics[width=\columnwidth]{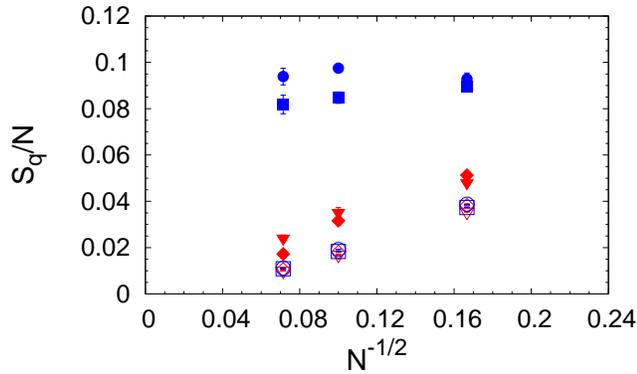}
\caption{\label{fig:sq}
(Color online) Size scaling of the spin-spin correlations $S_Q/N$ for 
$Q=(\pi,\pi)$. Data are for $t^\prime/t=0.6$ with $U/t=10$ (squares) and 
$U/t=20$ (circles), and $t^\prime/t=0.85$ with $U/t=10$ (triangles) and
$U/t=20$ (diamonds). Variational and fixed-node results are denoted by empty 
and full symbols, respectively. The variational results do not depend 
substantially upon $U$ and $t^\prime$. The fixed-node results indicate 
long-range order for $t^\prime/t=0.6$ but not for $t^\prime/t=0.85$.}
\end{figure}

\begin{figure}
\includegraphics[width=\columnwidth]{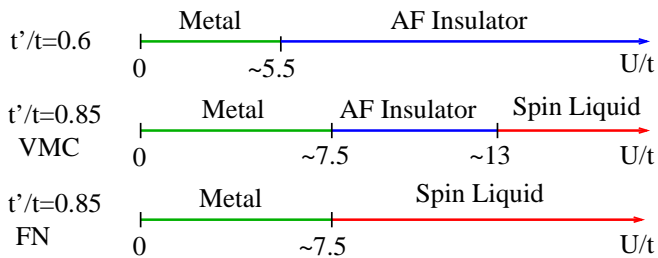}
\caption{\label{fig:phase_diagram}
(Color online) Proposed phase diagram for the two discussed hopping ratios, 
$t^\prime/t=0.6$ and $t^\prime/t=0.85$. In the first case, variational (VMC)
and fixed-node results (FN) indicate both a direct transition between a 
metal and an insulator with AF N\'eel order at a critical value of the 
electron-electron repulsion $U_{c}$. For $t^\prime/t=0.85$, the
variational results predict the existence of three different phases 
at increasing $U/t$: a metal, an AF insulator with N\'eel order and a 
spin liquid, while, within the fixed-node approximation, the non-magnetic 
ground state extends down to the metal-insulator transition.}
\end{figure}

\subsection{Insulating Phase}

In the insulating regime and for small frustrating ratios, the AF wave function
has always a lower energy than the spin-liquid state, and this fact is 
particularly evident close to the transition, see Fig.~\ref{fig:energy}. 
On the contrary, for the case with $t^\prime/t=0.85$, the BCS state competes 
with the AF one and it becomes better in energy for $U/t \gtrsim 13$, 
indicating an insulating spin-liquid behavior at large $U$ 
(notice that in this region $N_q \sim q^2$). In this regime, 
the BCS pairing is relevant, since the simple projected Fermi 
sea has a much higher energy, see Fig.~\ref{fig:energy}. 
Remarkably, the BCS and AF variational energies
are always quite close for $t^\prime/t=0.85$, suggesting that the actual 
ground state might be non-magnetic for all $U > U_c^{\rm MIT}$, or at least 
down to values lower than expected on the basis of the variational estimate. 
This fact is supported by the fixed-node calculations 
for the spin-spin correlations 
\begin{equation}
S_q = \frac{1}{N} \sum_{j,l} e^{i q (R_j-R_l)} \langle S^z_j S^z_l \rangle.
\end{equation}
In Fig.~\ref{fig:sq}, we report the size scaling of the variational and the 
fixed-node results by considering the BCS state as the {\it guiding function}. 
We stress the fact that, in the insulating regime, $S_q$ has a peak at the 
commensurate momentum $Q=(\pi,\pi)$. Remarkably, the fixed-node approach is 
able to recover a finite value of $S_Q/N$ for $Q=(\pi,\pi)$ (i.e., the square 
of the magnetic order parameter) in the thermodynamic limit for 
$t^\prime/t=0.6$, even though the BCS wave function is non magnetic. 
By contrast, $S_Q/N$ tends to zero for $t^\prime/t=0.85$ (both for $U/t=10$ 
and $20$), supporting the fact that the ground state is non magnetic for this 
frustrating ratio, even close to the metal-insulator transition. The resulting 
phase diagram is summarized in Fig.~\ref{fig:phase_diagram}.

\section{Discussion}\label{sec:conc}
We have studied the anisotropic triangular lattice at half filling away from 
the isotropic point $t^\prime=t$, with $t^\prime<t$, using both a 
Gutzwiller-Jastrow variational ansatz including backflow correlations as well 
as a Green's function Monte Carlo approach within the fixed node approximation.
We find that the square lattice states persist up to large values of 
$t^\prime/t<1$, both in terms of the d-wave superconducting order parameter 
as well as for the AF N\'eel ordering.

The main outcome of this work is that, thanks to the improvement given by 
backflow correlations, a spin-liquid wave function can be stabilized over 
magnetic states, for large but still moderate Coulomb repulsions and close 
to the isotropic limit. These variational results are corroborated by fixed-node
calculations. We find, in particular, that for $t^\prime/t=0.85$, which is 
relevant for $\kappa$-(ET)$_2$Cu$_2$(CN)$_3$,~\cite{nakamura,valenti} 
the insulating phase has a pure Mott character, without magnetic order. 
On the other hand, for $t^\prime/t=0.6$, suitable for 
$\kappa$-(ET)$_2$Cu$_2$(SCN)$_2$, (or even smaller $t^\prime/t$ values) 
the insulating phase always shows N\'eel order with $Q=(\pi,\pi)$. 

Let us finish by discussing our results also in comparison to other
calculations and experimental findings. First of all, in various papers, it
has been suggested that the spin-liquid phase can be stabilized by charge
fluctuations that may destabilize a magnetically ordered state. This claim has
been corroborated by calculations on Heisenberg models in presence of a 
ring-exchange term $J_4$ (that appears in the strong-coupling expansion in 
$t^4/U^3$).~\cite{motrunich} However, it turns out that the actual value of
$J_4$ for stabilizing a disordered phase is rather large and, probably, beyond
the validity of a perturbative expansion. The existence of a direct transition 
from a magnetic phase to a disordered one has been also found in the original
Hubbard model, by decreasing the on-site repulsion $U$.~\cite{clay,imada,kyung}
We do not find any evidence of such a possibility and, in our approach, the 
magnetic phase is stable in presence of charge fluctuations, even close to 
the metal-insulator transition: this is the case of $t^\prime/t=0.6$. 
Instead, the spin-liquid phase turns out to be directly connected with the one 
at strong coupling, while antiferromagnetic correlations become stronger when 
decreasing $U/t$. For example, for $t^\prime/t=0.85$, the variational state
with magnetic order has a slightly lower energy close to $U_c$ and we 
need to apply the Green's function Monte Carlo approach to extend the 
spin-liquid region down to the metal-insulator transition,
see Fig.~\ref{fig:phase_diagram}.
At this stage, we would also like to mention that the metallic phase is likely 
to be not superconducting. In fact, the BCS pairing $\Delta$ in the metallic 
region is slightly  suppressed when improving the accuracy of the variational 
wave function by considering backflow correlations and, moreover, it is reduced 
by a factor $2 \div 3$ when the lattice size is increased from $10 \times 10$ 
to $14\times 14$, see table~\ref{tab:delta}. This fact contrasts the previous 
claim of a possible superconducting phase close to the metal-insulator 
transition by Liu and collaborators.~\cite{nandini}

Another very important point is to clarify the nature of the low-energy
excitations. Very recently, thermodynamic measurements of the specific
heat suggested the possible existence of a Fermi surface of neutral, $S=1/2$
fermionic spinons.~\cite{yamashita} However, it should be noticed that such
a measurement involves a difficult subtraction of a divergent nuclear
specific heat, and instead the thermal conductivity (which is not affected by
a nuclear contribution) shows an activated behavior with a tiny gap of 
$0.46 K$.~\cite{yamashita2} This fact has been associated with the existence 
of spinless ``vison'' excitations.~\cite{sachdev} 
From our calculations, it appears that the disordered insulating phase cannot 
sustain a true spinon Fermi surface, as previously suggested both on 
variational calculations~\cite{motrunich} and field-theory 
approaches,~\cite{leelee} but it has Dirac points at $(\pm \pi/2,\pm \pi/2)$. 
In fact, the projected Fermi-sea has a much higher energy than our best
variational ansatz with BCS pairing, see Fig.~\ref{fig:energy}.
Should our results be correct, either a deeper investigation of the minimal 
microscopic model for describing organic charge-transfer salts is needed or
a reinterpretation of the experimental data is required.

We thank R. Valenti for very useful discussions. L.T. and C.G. thank partial 
support from the German Science Foundation through the Transregio 49.


\begin{thebibliography}{99}
\bibitem{notet} At half filling, the sign of $t^\prime$ is irrelevant.
\bibitem{noteph} Single-particle states are defined in the BCS Hamiltonian,
   after the transformation $c^\dagger_{i,\downarrow} \to c_{i,\downarrow}$.
\end{thebibliography}
\end{document}